\documentclass[prd, twocolumn, nofootinbib, floatfix, preprintnumbers]{revtex4-1}

\usepackage[mathscr]{euscript}
\usepackage{amsmath}
\usepackage{graphicx}
\usepackage{dcolumn}
\usepackage{bm}
\usepackage{epsfig}
\usepackage{amssymb,latexsym,mathrsfs}
\usepackage{graphicx}
\usepackage{color}
\usepackage{hyperref}
\usepackage{float}
\usepackage{diagbox}
\usepackage{textpos}

\hypersetup{
    colorlinks=true,
    linkcolor=red,
    citecolor=blue,
} 

\newcommand{\be}{\begin{equation}}
\newcommand{\ee}{\end{equation}}
\newcommand{\bs}{\begin{split}} 
\newcommand{\bea}{\begin{eqnarray}}
\newcommand{\eea}{\end{eqnarray}}

\newcommand{\kp}{\kappa}

\begin{document}

\title{Remnant-free Moving Mirror Model for Black Hole Radiation Field}
\preprint{MIT-CTP/5144}

\author{Michael R.R. Good${}^{1,2}$}
\author{Eric V.\ Linder${}^{2,3}$} 
\author{Frank Wilczek${}^{4,5,6,7}$}
\affiliation{${}^1$Physics Department, Nazarbayev University, Astana, Kazakhstan.\\
${}^2$Energetic Cosmos Laboratory, Nazarbayev University, Astana, Kazakhstan.\\ 
${}^3$Berkeley Center for Cosmological Physics \& Berkeley Lab, University of California, Berkeley, CA, USA.\\
${}^4$MIT, Cambridge, MA, USA.\\
${}^5$T. D. Lee Institute and Wilczek Quantum Center,
Shanghai Jiao Tong University, Shanghai, China.\\
${}^6$Arizona State University, Tempe, AZ, USA.\\
${}^7$Stockholm University, Stockholm, Sweden.}

\begin{abstract} 
We analyze the flow of energy and entropy emitted by a class of moving mirror trajectories which provide models for the radiation fields produced by black hole evaporation.  The mirror radiation fields provide natural, concrete examples of processes that follow thermal distributions for long periods, accompanied by transients which are brief and carry little net energy, yet they ultimately represent pure quantum states.   A burst of negative energy flux is a generic feature of these fields, but it need not be prominent.  
\end{abstract} 

\date{\today} 

\maketitle

%\tableofcontents

%%%%%%%%%%%%%%%%%%%%%%%%%%%%%%%%%%%%%%%%%%%%%%%%%%%%%%%%%%%%%%%%%%%%  
%\section{Nature Summary Paragraph}
%Requirements here: {\href{https://www.nature.com/nature/for-authors/formatting-guide}{Nature Formatting Guide}}\\

In the context of quantum field theory, moving mirror models consider the effect of imposing boundary conditions on a moving surface \cite{Davies:1976hi,purity,Fabbri:2005mw}.   There are prospects to realize interesting versions of these models experimentally \cite{mourou,wilson,wang} (and see below).  
Moving mirror models are thought to provide useful idealizations of black hole evaporation \cite{Hawking:1974sw}, though it remains unclear how far the analogy can be taken.  Here we analyze a class of moving mirror trajectories leading to radiation fields which have several properties that are remarkable in themselves, and desirable in a model of black hole evaporation:
\begin{itemize}
\item The emitted radiation follows a thermal (Planck) spectrum for arbitrarily long times.  This simulates the Hawking radiation process which, according to a semiclassical analysis, is thought to dominate the evaporation of an isolated black hole over most of its lifetime.  
\item The radiation field is limited in space and time.  
\item Apart from the radiation, the quantum fields approach their normal ground states (i.e.\ ``vacuum'') at early and late times. In this sense, there is no remnant \cite{horizonless,Chen:2014jwq,GTC,universe,MG15}.
\item Despite the thermal appearance of the bulk of the radiation field, the final state, including the radiation field, is a pure quantum state.  
\end{itemize}
This last point is especially interesting, for it embodies the ``information loss'' paradox and, within this class of models, resolves it.  This bundle of properties suggests that with use of these trajectories the moving mirror idealization of evaporating black hole radiation may be remarkably appropriate.  

Through a sum rule connecting the flows of energy and entanglement entropy, we show that these properties imply a period of negative energy flux.  The required flux need not be large, however, if it occurs at near-maximal entropy.   

%%%%%%%%%%%%%%%%%%%%%%%%%%%%%%%%%%% 
%\subsection{Reflection of a Black Mirror} 
\vspace{0.3cm} 

%{\it Planck Mirror---} 

{\it Model.} 
In moving mirror models we impose Dirichlet boundary conditions along the worldline of a ``mirror'' on quantum fields in 1+1 dimension. 
For concreteness we will focus on the case of a single massless scalar field, though most of our results generalize to general conformal field theories.
For our detailed analysis, we will focus on emissions accompanying a particularly simple and symmetric 2-parameter family of trajectories.  
As the analysis will make clear, the radiation fields emitted by moving mirrors following a wide variety of trajectories that share some broad qualitative features will share the good properties of this symmetric family.   

Our model trajectories are given by 
\be t(x) = -x -\frac{\sinh (2 \kappa  x)}{g}. \label{FTP}\ee
where $\kappa$ and $g$ are free parameters.   They are inspired by the ``black mirror'' trajectory \cite{MG14one,MG14two,Good:2016MRB}
\be t(x) = -x -\frac{e^{2 \kappa  x}}{\kappa }. \label{BHC}\ee
The black mirror evolves to an asymptotic light-like trajectory \cite{spin,Good:2013lca} of an eternal black hole horizon \cite{Carlitz:1986nh,paper2}.  It emits an exactly Planckian spectrum of radiation with temperature $T = \frac{\kappa}{2\pi}$.  Our trajectories are $PT$ symmetric versions of the black mirror, in which we also introduce an additional parameter.  

%%%%%%%%%%%%%%%%%%%%%%%%%%%%%%%%%% 
%\subsection{Trajectory Dynamics}

\begin{figure}[ht]
\centering 
\includegraphics[width=3.2in]{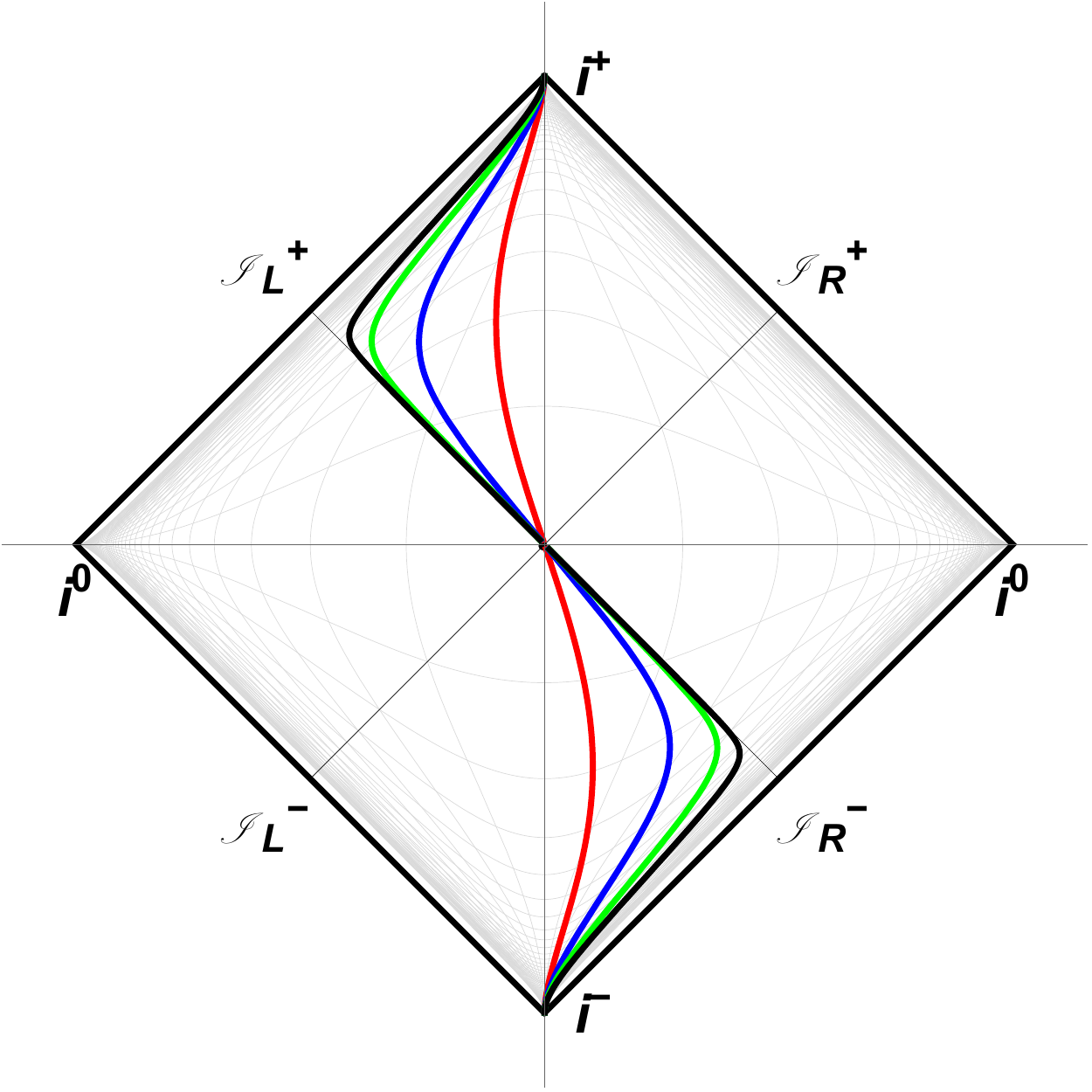} 
\caption{The trajectory, Eq.~(\ref{FTP}), plotted in a conformal diagram. Here $\kappa = 1$ and $g = 10^n$.  Red, Blue, Green, Black are $n=0,1,2,3$, respectively. 
%This color scheme will be abandoned in subsequent figures. 
} 
\label{fig:Penrose_Plot} 
\end{figure}

Fig.~(\ref{fig:Penrose_Plot}) displays trajectories defined by Eq.~(\ref{FTP}) within a Penrose 
conformal diagram. Note that 
for large $t$ the mirror becomes static, $\dot{x} \rightarrow -\frac{1}{\kappa t}$: that is, $t(x)$ approaches a vertical asymptote as $x\rightarrow \pm\infty$.  Thus this trajectory displays asymptotically static boundaries \cite{walkerdavies,Good:2013lca,paper1,paper3}.   Near $t=0$, on the other hand, the mirror bends toward a null trajectory, where its 
velocity approaches the speed of light.  The mirror trajectory simulates the effect of dynamical geometry in generating radiation fields.  When static, i.e. at early and late times, it plays the role of the center of radial coordinates, but at intermediate times it simulates the mathematics of the black hole horizon in Hawking's calculation.  

Consistent with the ``no remnant'' interpretation, the static initial and final states represent empty space.
More precisely, we have the velocity 
\be 
\frac{dx}{dt} ~\equiv~ V(x)=-\frac{g}{g+2 \kappa  \cosh (2 \kappa  x)}\ . 
\ee
It is zero in the limit $x\rightarrow \pm \infty$, and 
has its maximum absolute value at $x=0$, where $V_\textrm{max}=-g/(g+2\kappa)$.  For $g\gg\kappa$, $V_\textrm{max}\to-1$.   At late times the velocity scales as inverse time.   Small and innocuous modifications of the trajectory at late (and early) times could bring the mirror strictly to rest finally (and initially).   To do this one can employ smooth but non-analytic functions, as are used to construct smooth partitions of unity \cite{tu}.

%%%%%%%%%%%%% 

%%%%%%%%%%%%%%%%%%%%%%%%%%%%%

%%%%%%%%%%%%%%%%%%%%%%%%%%%
%\subsection{Energy and Entropy to $\mathscr{I}^+_R$} 

%\bigskip

{\it Radiation Field.} 
The expectation value of the stress tensor can be evaluated analytically, in terms of the mirror trajectory $t(x)$.   It is straightforward to evaluate the energy flux $F(x)$ at the mirror, as derived from the stress tensor \cite{Davies:1976hi}, at right null infinity. 

\begin{figure}[ht]
\centering 
\includegraphics[width=3.2in]{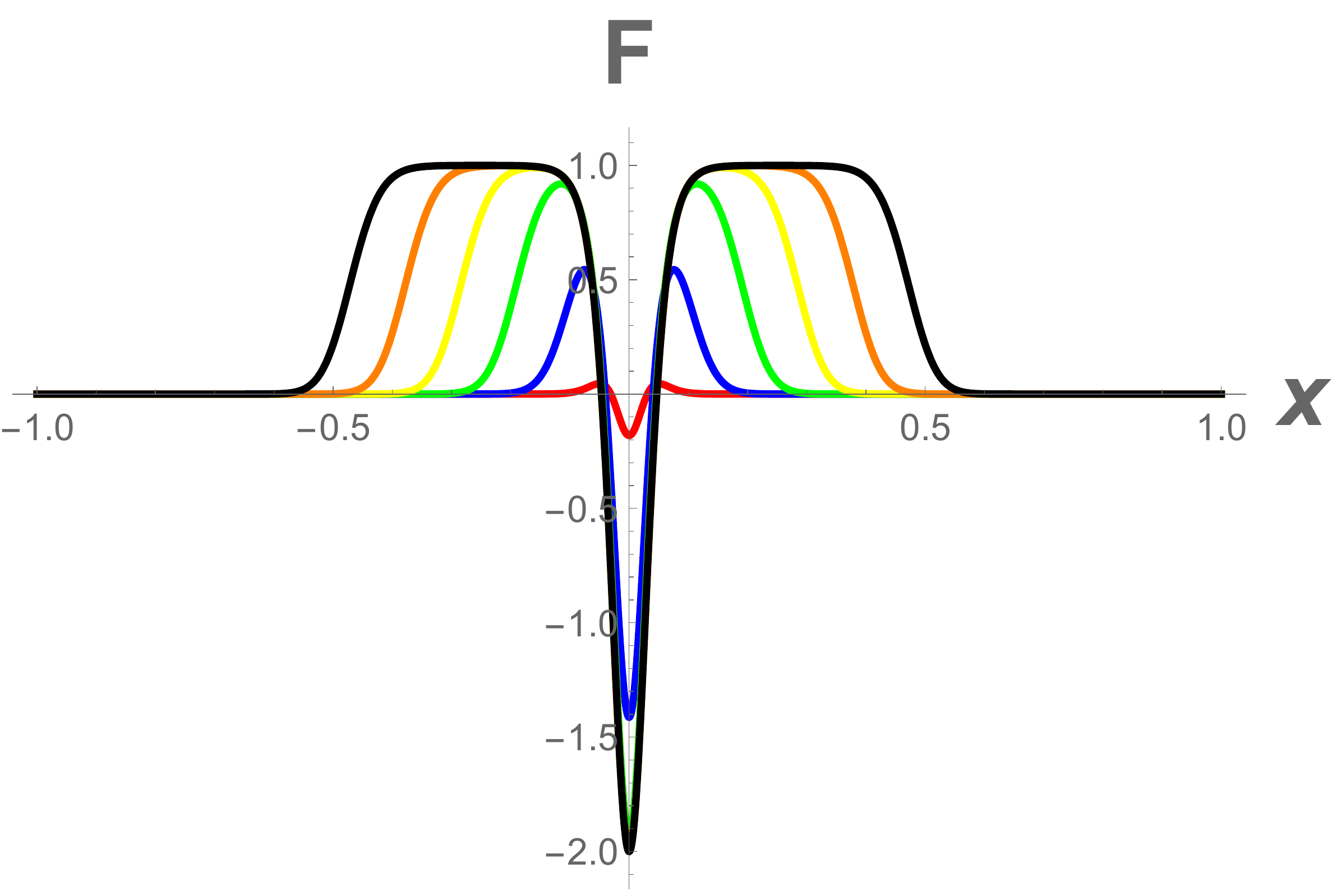} 
\caption{The energy flux from the symmetric mirror is plotted with $\kappa = \sqrt{48\pi} = 12.279$ so that thermal equilibrium is seen to be $F(x) = 1$. We show $g = 10^n$ where Red, Blue, Green, Yellow, Orange, Black is $n = 1,2,3,4,5,6$, respectively. Note the color scheme differs from Fig.~(\ref{fig:Penrose_Plot}) but is consistent with all 
other figures. 
} 
\label{fig:Energy_Flux} 
\end{figure}

As can been seen in Fig.~(\ref{fig:Energy_Flux}), when $g/\kappa \gg 1$ we have two long lived plateaus.  These plateaus occur at the flux value
\be F_{\rm thermal} = \frac{\kappa^2}{48\pi},\label{flux_plateau_48pi}\ee
which is the energy flux value associated with thermal emission in the analog black mirror.  To leading order in $\kappa/g\ll1$ the normalized flux approaches
\be F(x)/F_{\rm thermal} = 1-3\,\text{sech}^2(2\kappa x) + \mathcal{O}(\kappa/g)\ .\label{energy_flux}
\ee 
A striking feature and perhaps surprising feature is the burst of negative energy flux around $x=0$.  It saturates to twice the height of the plateau, and finite width as the plateaus expand. We will demonstrate below, using the connection between energy flow and geometric entropy, that negative energy flux is a necessary feature of the radiation fields associated with remnant-free moving mirror models, following from unitarity.

The total energy flux observed at right null-infinity is finite.  For large $g\gg\kappa$ we have 
%\be E = \left(2\ln \left[\frac{2 g}{\kappa}\right]-5\right)\frac{\kappa}{48\pi}\ ,\label{totalenergy}\ee
\be E = \frac{\kappa}{24\pi}\ln \frac{g}{\kappa}\ . \label{totalenergy}\ee
The moving mirror model contains no variable that corresponds directly to the black hole mass, but one can define an effective mass parameter by imposing the semiclassical (Hawking) relationship between the radiation flux and mass. For our trajectories the effective mass, so defined, remains constant on the long plateaus.  Thus they do not reflect the expected increase of temperature (associated with decrease of black hole mass).   It should be possible to construct approximately ``self-consistent'' trajectories which incorporate that feature, but we will not attempt that here.   Here we only note that we are free to choose $g$ in such a way that the energy $E$ corresponds to the mass of the black hole which emits radiation at temperature $T$.  This entails $\ln(g/\kappa) \sim M^2$, with $M$ measured in Planck units.  Thus $g/\kappa \gg 1$ is appropriate in modeling semiclassical black holes.  

%\bigskip

% \cite{Fabbri:2005mw}.  {\bf what does this paper do?}

%\bigskip

%%%%%%%%%%%%% 

\begin{figure}[ht]
\centering
\includegraphics[width=3.2in]{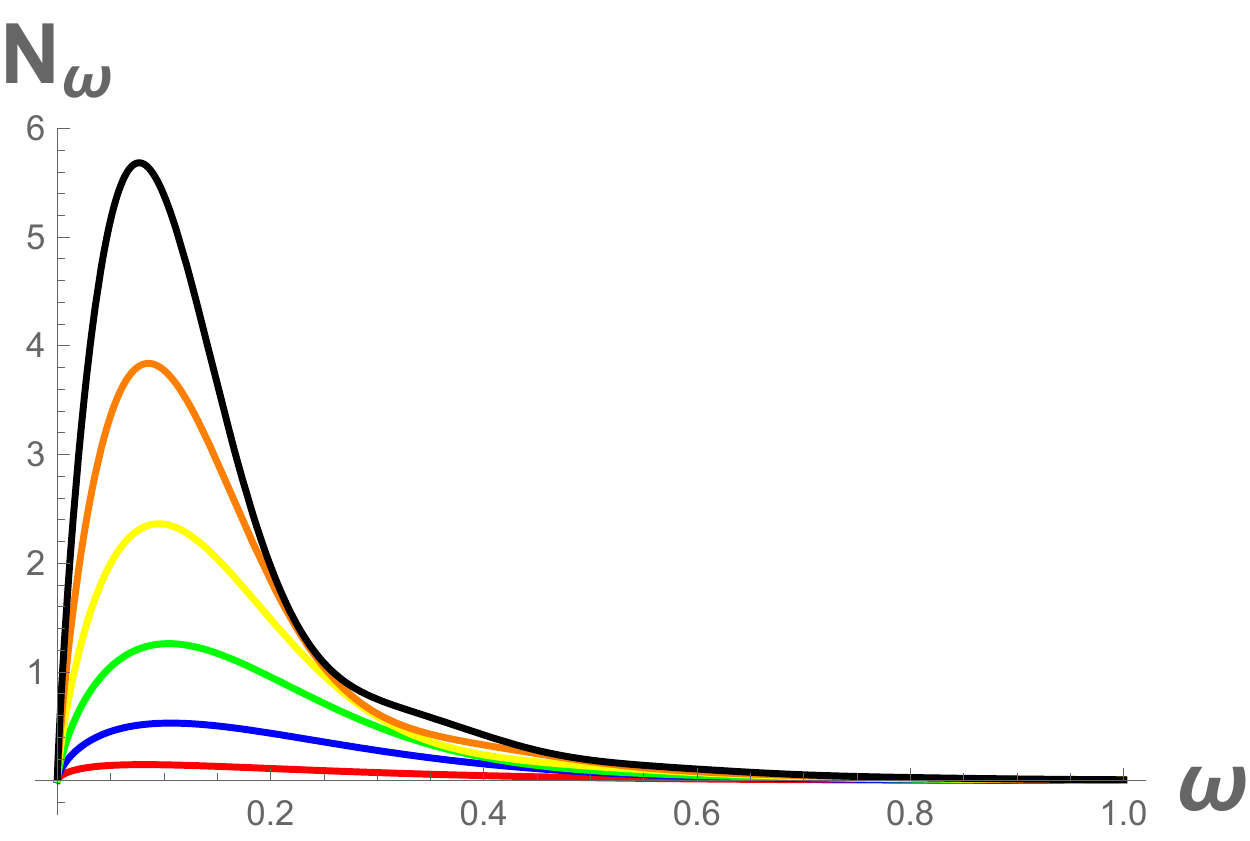} 
\caption{The spectrum, $N_\omega$, Eq.~(\ref{N_w}), or particle count per mode detected at right null infinity, of the 
asymptotically static thermal mirror is plotted vs frequency $\omega$, 
in units of $\kappa$.  
%The different colored lines correspond to different $g$ values: $g =  10^n$, where $n = 1,2,3,4,5,6$ for Red, Blue, Green, Yellow, Orange, Black, respectively.  
This asymptotically static solution shows no infrared divergence in the number of soft particles. 
\label{fig:Particle_Spectrum}} 
\end{figure}

The central quantities for calculating 
particle production are the beta Bogolyubov coefficients, 
which can be calculated analytically. 
%, with  
%\be \beta_R(\omega,\omega') = -\frac{\sqrt{\omega\omega'}}{\pi  \kappa  \omega_p }e^{-\frac{\pi  \omega }{2 \kappa }} K_{\frac{i \omega }{\kappa }}\left(\frac{\omega_p}{g}\right) ,\label{beta}\ee
The particle spectrum per mode per mode is
\be |\beta|^2 = \frac{\omega \omega' e^{-\frac{\pi  \omega }{\kappa }}}{\pi ^2 \kappa ^2 \omega_p^2}\,\left|K_{\frac{i \omega }{\kappa }}\left(\frac{\omega_p}{g}\right)\right|^2\ ,\label{rightspec1}\ee
where $K_n(z)$ is a modified Bessel function of the second kind and $\omega_p \equiv \omega'+\omega$, the sum of the 
in and outgoing mode frequencies. The spectrum, $N_\omega$, or particle count per mode, detected at right null infinity surface  $\mathscr{I}^{+}_R$ is found by integrating Eq.~(\ref{rightspec1}), 
\be N_\omega = \int_0^\infty  |\beta|^2 \; d \omega'\ . \label{N_w}\ee
Fig.~(\ref{fig:Particle_Spectrum}) illustrates the results for different $g$ values.   For large $g$ we approach a Planck spectrum, but not uniformly.  There is always a zero at strictly 0 frequency, and the total number of radiated particles is always finite.  The total energy can be retrieved using the particles as a sum over quanta, 
\be E = \int_0^\infty \int_0^\infty  \omega \cdot |\beta|^2 \; d \omega \; d \omega'\ , \ee
This consistent global energy result via Eq.~(\ref{rightspec1}) is the same energy as  Eq.~(\ref{totalenergy}), which was derived earlier by integrating the local stress tensor. 

\begin{figure}[ht]
\centering
\includegraphics[width=3.2in]{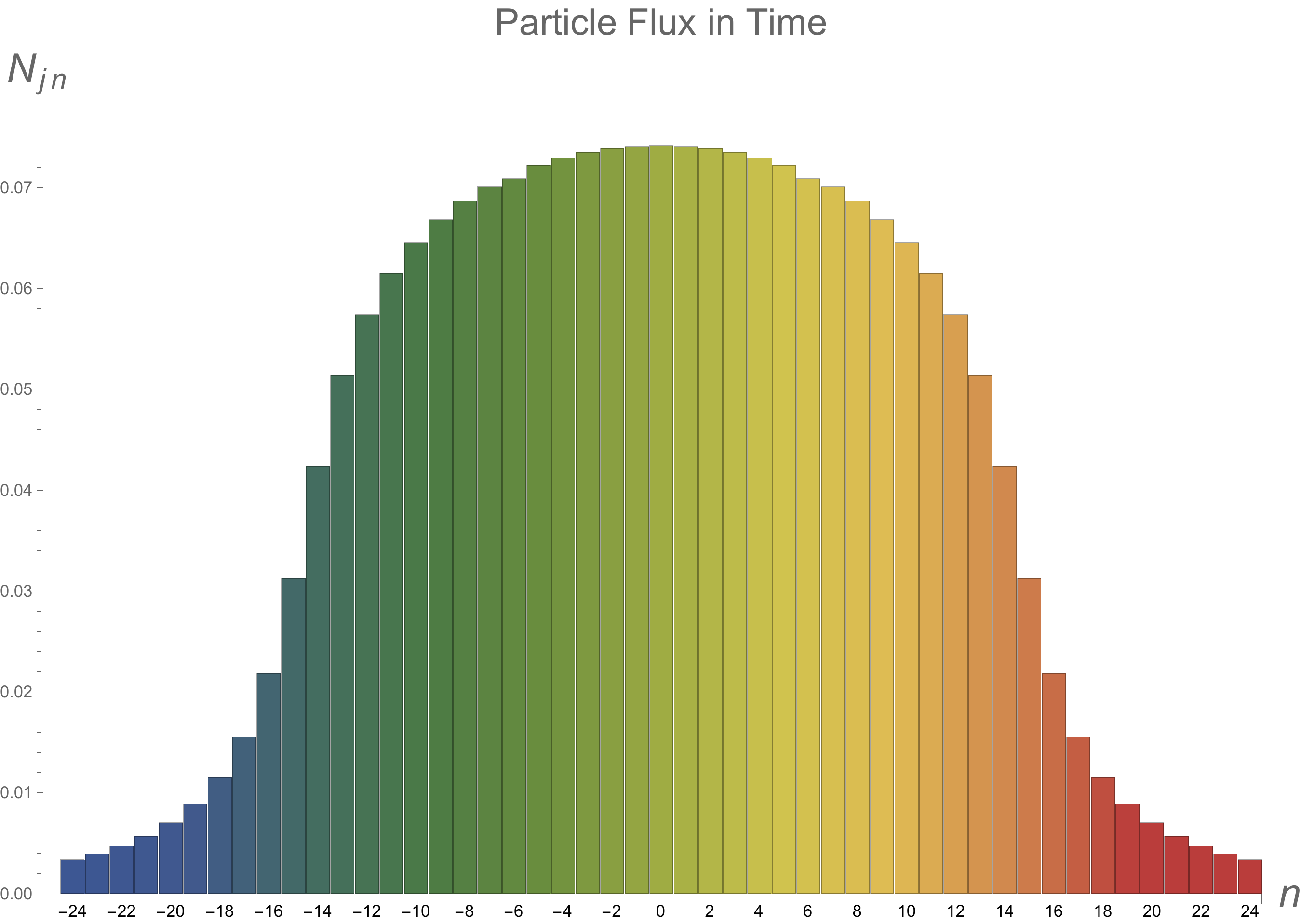} 
\caption{The discrete spectrum, $N_{jn}$, time evolved.  Here the system is set with $\kappa=1$, and $g=10^{10}$.  The detector is set with $j=0$, $n=(-24,24)$, $\epsilon = 4$.  %The expected total particle count of this system is $N=2.209$. 
%The sum of the displayed columns is $N=2.141$. Inclusion of more time bins (`$n$') will yield the total count.  
Notice the flattened plateau centered around $n=0$.  
\label{fig:Particle_Flux_In_Time}}
\end{figure}

For large $g/\kappa \gg 1$ the integrand $|\beta|^2\equiv N_{\omega\omega'}$  then has the simple form, to lowest order and assuming a true Cauchy principal value, 
\be N_{\omega \omega'} =\frac{e^{-\omega \pi} \omega' \text{csch}(\pi  \omega)}{2 \pi  \left(\omega'+\omega\right)^2} \approx \frac{1}{\pi \omega' }\frac{1}{ e^{2 \pi  \omega}-1},\ee
in units of $\kappa$,
where the last step simplifies the prefactor by considering the large frequency regime, $\omega'\gg \omega$.  The appearance of the Planck distribution for large $g$ is consistent with the constant plateau energy flux of Eq.~(\ref{flux_plateau_48pi}) for $g \gg \kappa$ as seen in Fig.~(\ref{fig:Energy_Flux}) and the flattening plateau for constant particle emission over time in Fig.~(\ref{fig:Particle_Flux_In_Time}).  It parallels Hawking's calculation of radiation from a fixed black hole background.

%%%%%%%%%%%%%%%%%%%% 
%\subsection{Large Production of Particles}
For any finite value of $r \equiv g/\kappa$ the total number of particles created is finite, though it increases without limit as $r$ grows.  
%with this model by direct (single) numerical integration of the analytic answer of Eq.~(\ref{N_w}),
%\be
%N = \int_0^\infty N_{\omega} \; d\omega\ , \label{totalparticles}
%\ee
%when the dimensionless ratio  is very large.  This gives the most particle production.
%\footnote{Instead of using the more intuitive large maximum speed $V_{\textrm{max}}$ or large maximum rapidity, $\eta$, as done in Eq.~(\ref{Nofeta}), it is much %more numerically tractable to compute with large $r \equiv g/\kappa$.} 
With $r=10^{10^7}$ we find $N\approx 10^7$ particles.  The absence of an ``infrared catastrophe'' in the number of soft quanta reflects the absence of a physical remnant.

%%%%%%%%%%%%%%%%%%%%%%%%%%%%%%% 

%%%%%%%%%%%%%%%%%%%% 

%%%%%%%%%%%%%%%%%%%%%%%%%%%%%
%\subsection{Total Energy to $\mathscr{I}^+_R$}

%%%%%%%%%%%%%%%%%%%%%%%%%% 
%\subsection{Entropy Flux at $\mathscr{I}^+_R$} 

%\bigskip

\begin{figure}[ht]
\centering 
\includegraphics[width=3.2in]{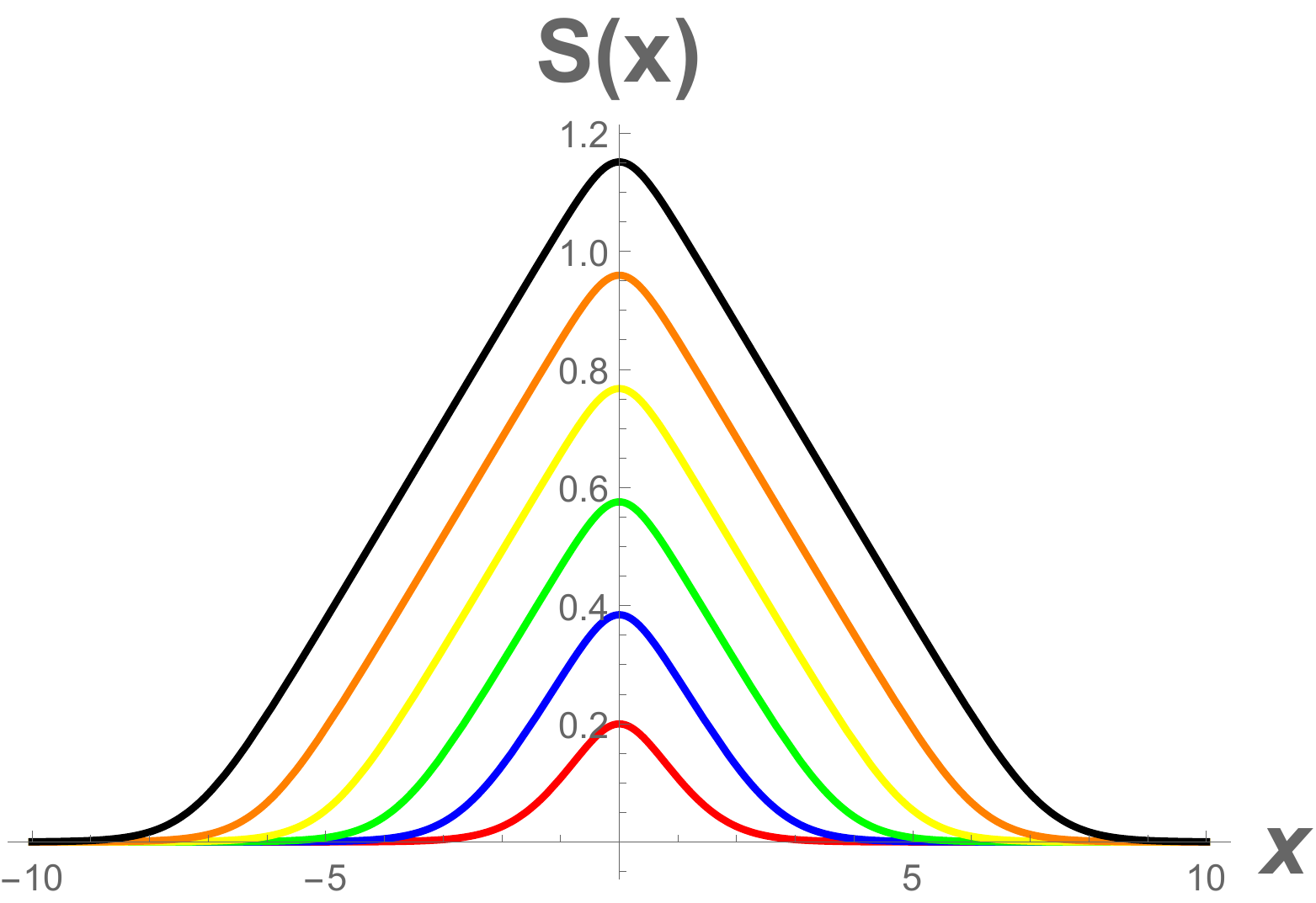} 
\caption{The asymptotically static mirror with finite energy and finite particle count has entropy flux, Eq.~(\ref{S(x)}), plotted with different $g$ values (same color scheme as Fig.~\ref{fig:Energy_Flux}).} 
%The different colored lines correspond to $g =  10^n$, where $n = 1,2,3,4,5, 6$ for Red, Blue, Green, Yellow, Orange, Black, respectively. 
%The Gray line corresponds to $g=1$, while $\kappa=1$ throughout. 
\label{fig:Entropy_Flux} 
\end{figure}

{\it Entropies.} 
Both thermodynamic entropy and entanglement entropy play an important role in particle creation models \cite{Holzhey:1994we}.  For our purposes, the most enlightening measure of entanglement (see also harvesting \cite{harvest}) is the renormalized entanglement entropy of the state at future null infinity to the left (or right) of a given value of null time $u$.  Heuristically, this represents a flow of entanglement entropy through $u$.   In the moving mirror model, it has the simple form  $-6S = \eta$, where $\eta$ is the rapidity of the mirror as it crosses $u$.   In our model we have a simple expression for $S$ expressed as a function of $x$, the corresponding position of the mirror:
%\be S(x) = -\frac{1}{6} \tanh^{-1} t'(x)^{-1}\ ,\ee 
%radiated to the right is symmetric about its maximum at $x=0$, with  
\be 
S(x)% &=&  \frac{1}{6}\tanh ^{-1}\left(\frac{g}{g+2 \kappa  \cosh (2 \kappa  x)}\right)  \label{S(x)}\\ 
%&=&
=\frac{1}{12}\ln\left(1+\frac{g}{\kappa\cosh(2\kappa x)}\right) \ .\label{S(x)} 
\ee 
%The later 
The entropy vanishes for the asymptotic spatial positions, as it should since the evolution is unitary and entails no information loss.  Fig.~(\ref{fig:Entropy_Flux}) illustrates this entropy flux.   %%%%%%%%%% 

The energy flux $F$ is related analytically to the entropy $S$ according to (e.g. \cite{paper3,entropyevolution,bsmer}) 
\begin{eqnarray}
F(u) ~&=&~ \frac{1}{2\pi} (6 S^{\prime 2} + S^{\prime \prime}) \\
~&=&~ \frac{1}{2\pi} e^{-6S}\, \frac{d}{du} \left(e^{6S} \frac{d}{du}S\right)\ . \label{F_as_derivative}
\end{eqnarray}
From Eq.~(\ref{F_as_derivative}) and the assumption that $S$ becomes constant for $u \rightarrow \pm \infty$ we derive the sum rule, 
\be\label{sum_rule}
\int\limits^\infty_{-\infty} du \, e^{6S(u)}  F(u) ~=~ 0\ . 
\ee
Thus, on general principles, the flux $F(u)$ will have a negative region for the radiation field of a remnant-free pure state.   Note that for purposes of fulfilling the sum rule the negative flux has greatest leverage when it occurs at the maximal $S(u)$, 
as we see realized in Figs.~(\ref{fig:Energy_Flux}) and (\ref{fig:Entropy_Flux}).  The flux of statistical mechanical entropy associated with the thermally-distributed flux on the plateau tracks $S^\prime$ for $x > 0$ and $-S^\prime$ for $x<0$.  
%\vspace{0.3cm} 

%%%%%%%%%%%%%%%%%%%%%%%%% 
%\subsection{Time Dependence of Particle Count} 

{\it Particle Counts in Time.} 
Time evolution can be resolved with the use of wave packets \cite{Hawking:1974sw} $\beta^\epsilon_{jn}$ that pick out frequencies near $j\epsilon$ at retarded time $u = 2\pi n \epsilon^{-1}$ with width $2\pi \epsilon^{-1}$.  
This localization of the global beta coefficients corresponds to the sensitivity response of a particle detector at a given time, frequency, and bandwidth 
%This localization of the global beta coefficients inspects the response of a particle detector with sensitivity to the expected time-frequency uncertainty in duration band-width 
\cite{Good:2013lca,Anderson:2019njv}.  With large $g\gg\kappa$, and good time resolution (i.e., large $\epsilon$; the particles pile up in the single $j=0$ bin) one observes a flattening plateau, Fig.~(\ref{fig:Particle_Flux_In_Time}).  Remarkably, the plateau extends through the $n=0$ time bin, with no obvious scar from accompanying the negative energy flux.  
%In earlier moving mirror solutions \cite{signatures}, where particle emission did display non-thermal and auxiliary signatures of energy flux.  

%%%%%%%%%%%%%%%%%%%%%%%% 

%The thermal nature of the particle emission can be 
%established from a 
%Defining dimensionless quantities $a \equiv \omega/\kappa$ and $a' \equiv \omega'/\kappa$, one can write the total count as
%  \be 
% N=\frac{1}{\pi^2}\int da\,da'e^{-\pi a}\frac{aa'}{(a+a')^2} \left|K_{ia}\left(\frac{a+a'}{g/\kappa}\right)\right|^2\,. \label{totalparticles} 
% \ee 
%closed-form analytic result for the spectrum, Eq.~(\ref{N_w}), with $|\beta|^2 \equiv N_{\omega\omega'}$ 
%given by Eq.~(\ref{rightspec1}). 
%\be N_\omega = \int_0^\infty N_{\omega\omega'} d\omega', \label{N_w}\ee
%where, using dimensionless frequencies in units of $\kappa$, 
%\be N_{\omega \omega'} \equiv \frac{e^{-\pi \omega}}{\pi^2}\frac{\omega\omega'}{(\omega+\omega')^2} \left|K_{i\omega}\left(\frac{\omega+\omega'}{g/\kappa}\right)\right|^2\,, \ee 
%but its appearance is complicated.  Regardless, it is the first finite
%analytic expression for the particle spectrum, $N_\omega$, of any particle 
%creation model, whether cosmological, black hole, or 
%accelerating horizon (mirror). 

%This also allows the total particle count to be obtained by 
%a single numerical integration rather than a double 
%numerical integration over the Bogolyubov coefficients. 

%%%%%%%%%%%%%%%%%%%%%%%%%%%%% 
%%%%%%%%%%%%%%%%%%%%%%%%%%%%% 
%%%%%%%%%%%%%%%%%%%%%%%%%%%%%%% 
%%%%%%%%%%%%%%%%%%%%%%%%%%%%%%% 
%%%%%%%%%%%%%%%%%%%%%%%%%%%%% 
%%%%%%%%%%%%%%%%%%%%%%%%%%%%% 
%%%%%%%%%%%%%%%%%%%%%%%%%%%%%%% 
%%%%%%%%%%%%%%%%%%%%%%%%%%%%%%% 
%%%%%%%%%%%%%%%%%%%%%%%%%%%%% 
%%%%%%%%%%%%%%%%%%%%%%%%%%%%% 
%%%%%%%%%%%%%%%%%%%%%%%%%%%%%%% 
%%%%%%%%%%%%%%%%%%%%%%%%%%%%%%% 

%\vspace{0.3cm} 

{\it Summary and Discussion.} 
The moving mirror models discussed here produce radiation fields that look thermal for long periods of time, and limited transients, yet represent pure states.  This, together with their geometric character, suggest that they may provide instructive models for quantum evaporation of black holes, which plausibly -- yet paradoxically -- have those properties.   

A striking feature, here simply shown to be generic, of remnant-free moving mirror models is the occurrence of negative energy flux.  Although states with locally negative values of energy density are known to occur in several contexts, including the intensely studied Casimir effect, their occurrence is somewhat unusual and there seems to be no general understanding of their properties.  In our context, the derivation of the sum rule Eq.\,(\ref{sum_rule}) connects negative energy flux to the purity of the quantum radiation field, though it does not provide a mechanistic explanation of the connection. 

Within the specific models analyzed here, the independent signatures of the negative energy flux appear to be subtle.  Specifically, as a fraction of the overall process  it is strictly limited in time and energy (for 
$g\gg\kappa$, $E_{\rm NEF}=-(\kp/24\pi)[\sqrt{6}-\tanh^{-1}\sqrt{2/3}]\approx -0.017\kp$), and it does not appear prominently in the response of quasi-realistic particle detectors, Fig.~(\ref{fig:Particle_Flux_In_Time}), unlike many positive energy flux signatures \cite{signatures}.  

In order to satisfy the sum rule Eq.\,(\ref{sum_rule}) with an inconspicuous negative energy flux, it is important that the flux occur where the entanglement entropy $S$ is large.  If this possibility is to be relevant to physical black hole evaporation, it should therefore act early in the black hole's history, or at regular intervals throughout.  To our knowledge no existing semiclassical treatment of black hole evaporation, including ones which attempt to incorporate back-reaction, contains such effects.  It is conceivable that better approximations, either within general relativity itself or in the larger framework of string theory, could display them.  It is tempting to speculate, in view of the connection to entropy, that entropic forces come into play, modifying the space-time geometry.   As we have seen, the required effects may not need to be large.   

Independent of their possible connection to black holes, the unusual features predicted to occur in radiation fields produced by moving mirror models are interesting in themselves, especially as they confront the tension between quantum purity and apparent thermality.   The availability of arrays containing very large numbers of mirrors whose orientation can be programmed flexibly might offer another road (in addition to \cite{mourou,wilson,wang}) to realizing such models.

%%%%%%%%%%%%%%%%%%%%% 
\acknowledgments 

MG is funded by the ORAU FY2018-SGP-1-STMM Faculty Development Competitive Research Grant No. 090118FD5350 at Nazarbayev University,  and the state-targeted program ``Center of Excellence for Fundamental and Applied Physics" (BR05236454) by the Ministry of Education and Science of the Republic of Kazakhstan. 
EL is supported in part by the Energetic Cosmos Laboratory and by 
the U.S.\ Department of Energy, Office of Science, Office of High Energy 
Physics, under Award DE-SC-0007867 and contract no.\ DE-AC02-05CH11231.  FW's work is supported by the U.S. Department of Energy under grant Contract  Number DE-SC-0012567, by the European 
Research Council under grant 742104, and by the Swedish Research Council under Contract No. 335-2014-7424.

%%%%%%%%%%%%%%%%%%%%%%%%%%%%%%%%%%%%% References in order

\end{document}